\documentclass[pre,showpacs]{revtex4}
\usepackage{amsmath}
\usepackage{latexsym}
\usepackage{epsfig}

\begin{document}
\title{Actin polymerization front propagation in a comb-reaction system}
\author{A. Iomin$^{1}$, V. Zaburdaev$^2$, T. Pfohl$^3$}
\affiliation{$^1$Department of Physics, Technion, Haifa 32000,
Israel \\
$^2$ Max-Planck-Institute for the Physics of Complex Systems,
N\"uthnitzer Str. 38 D-01187, Dresden, Germany \\
$^3$ Department of Chemistry, University of Basel, Klingelbergstrasse 80
CH-4056 Basel, Switzerland
}
\date{\today}

\begin{abstract}
Anomalous transport and reaction dynamics are considered by providing
the theoretical grounds for the possible experimental realization of actin polymerization in
comb-like geometry. Two limiting regimes are recovered, depending on the concentration
of reagents (magnesium and actin). These are both the failure of the reaction front propagation and a finite speed corresponding to the Fisher-KPP long time asymptotic regime.
\end{abstract}

\pacs{05.40.Fb, 82.40.-g}

\maketitle

\section{Introduction}

Microfluidics is an indispensable tool of modern bio-physical
research. It allows to perform complex single-cell experiments
with an immense throughput and high level of control. A flexible
design allows for custom geometries and control of flows and
chemical reactions. Recently, to probe the dynamics of actin
polymerization, as well as 
to use the geometry of microfluidic device having the main supply channel with
numerous identical side channels or chambers of different shapes
The main channel serves to deliver and fill the
side chambers with reagents where the corresponding reacting can
be and observed. The flow in the main channel and diffusion in the
side-channels are dominating means of transport in such devices.
Remarkably, the process of diffusion in this particular geometry
was extensively studied in the context of anomalous diffusion. It
is known as a comb model and it was demonstrated that the
transport of particles along the main channel (called backbone in
the model) can become subdiffusive when the particles get trapped
by diffusing into the side channels. Until recently it was mostly
an abstract model, which was, however, extremely useful in
understanding the principles of anomalous subdiffusive transport
(many references here). The experimental setup on actin
polymerization \cite{pfohl1,pfohl2} is the direct implementation of the
comb model \cite{white,weiss,havlin}, where the effects of complex diffusion should have a
substantial effect on the observed phenomena. Interestingly, the
comb structure not only leads to an anomaly in transport but also
to a very remarkable effects on the propagation of chemical
reactions \cite{iomend}.

The goal of this paper is to combine the consideration of
anomalous transport and reaction dynamics to provide the
theoretical grounds for the corresponding experimental efforts.
Our analytical results on reaction propagation can help to guide
the design of microfluidic devices but also can lead to real
experimental tests of anomalous diffusion and reaction dynamics.
For the reaction of polymerization, depending on the concentration
of reagents we can recover such remarkable phenomena as the
failure of reaction front propagation \cite{sokolov,iomsok} or a finite speed or a
Fisher-KPP long time asymptotic regime \cite{kpp,fisher}.

\section{Mapping of the Laplace operator on a Comb equation}

Mapping of the Laplace operators in the $3D$ complicated continuous-discrete geometry on a continuous comb model equation, relates to averaging over $xyz$-space.
In reality, we have two Laplace operators acting on distribution functions in a bulk $P_b(x,y,z)$ and in fingers (where reactions take place) $P_f(x,y,z)$.

(\textit{i}) In the bulk of infinite length along the $x$ coordinate and  $yz$ surface cross-section $a\times a$ we have
for the Laplace operator
\[D\Delta P_b(x,y,z)=D(\partial_x^2+\partial_y^2+\partial_z^2)P_b(x,y,z)\]
with the diffusivity of magnesium $D$ and the boundary conditions
\[\partial_xP|_{x=0}=\partial_xP|_{x=a}=\partial_yP|_{y=0}=\partial_yP|_{y=a}=0\, ,\]
the following algorithm of mapping can be suggested.

Integration over $z$ leads to the disappearance of the $z$ component due to the boundary condition.
Integration over $y$ in the bulk yields zero except those $y$ regions where the bulk is connected with the fingers. Plunging the fingers inside the bulk, one obtains that this region is at $y=0$.
Therefore, we have
\[D\Delta  P_b(x,y,z)\Rightarrow aD\partial_x^2P(x,y=0)\, .\]

(\textit{ii}) The fingers have height $h$ (along the $y$) and the $xz$ cross-section is of size $b\times b$. Therefore,
we have the following Laplace operator with diffusivity $d$ inside the fingers
\[d\Delta P_f(x,y,z)=d(\partial_x^2+\partial_y^2+\partial_z^2)P_f(x,y,z)\, .\]

Integration/averaging over $x$ and $z$ leads to zero for everywhere, except for $\partial_y^2$ in periodic regions
\[\partial_y^2P(x,y)\sum_n\delta(x-nL)\Rightarrow \frac{b}{L}\partial_y^2P(x,y)\, ,\]
where $L$ is a distance between the fingers.
Combining both equations, one obtains a kinetic equation for the magnesium transport along the comb structure
with parameters related to experimental geometry $a,b,L,h$ and diffusion coefficients $D$ and $d$
\begin{equation}\label{comb}
\partial_tP=D\delta(y/h)\partial_x^2P+\rho d\partial_y^2P\, .
\end{equation}
Here $\rho=bh/La$ is a dimensionless density of the fingers. In what follows we will use the notation for the coefficient
\[d_{\rho}=d\cdot\rho\, .\]
However, the scaling of the $\delta$ function by $h$ can be inconvenient in the limit $h\rightarrow\infty$. Therefore, in what necessary, we scale the $\delta$ function by
\begin{equation}\label{hb}
h_b=\sqrt{hb}\, ,~~~~~~~~~~ \delta(y/h)\rightarrow \delta(y/h_b)\, .
\end{equation}
In this case,
we have $h\rightarrow\infty$ and $b\rightarrow 0$, and in these limits $hb={\rm const}$. This scaling supports the density $\rho$. Say $hb=aL$ and in this subdiffusive limit $\rho=1$.

The boundary conditions are
\[P(x=\pm\infty,y,t)=\partial_xP(x=\pm\infty,y,t)=0~\mbox{and}~ \partial_yP(x,y=\pm h,t)=0\, .\]
The initial condition is
\[P(x,y,t=0)=P_0(x)\delta(y)\, .\]

Amending comb equation (\ref{comb}) with reaction term inside fingers $\rho\cdot C(P)$, we have
\begin{equation}\label{comb_r}
\partial_tP=D\delta(y/h_b)\partial_x^2P+ d_{\rho}\partial_y^2P-\rho C(P)\, ,
\end{equation}

Probably, another mapping on the comb model can be made in the framework of a Dykhne model
\cite{dkm2004,dvork2009,dvorkm2010}. In this case, dimensionless density is $\rho=\frac{b^2h}{a^2L}$.
In some extend, this density sounds more physically. But this leads to additional argumentation of limits, for example
$h\rightarrow \infty$ and $b\rightarrow 0$. Namely $hb^2\rightarrow a^2L$ and
$h_b\rightarrow h_{bb}=[hb^2]^{\frac{1}{3}}$.

\section{Few points on reaction for definition of the reaction term}

For the reaction-polymerization, we use the following \textit{stoichiometry} expression
\[ A+B \Rightarrow C\, ,\]
where
\begin{eqnarray}
& [A] & ~ \mbox{is concentration of actin} \nonumber \\
& [B] &  ~ \mbox{is concentration of magnesium}  \nonumber \\
& [C]  &~ \mbox{is concentration of polymer}  \nonumber
\end{eqnarray}

In general case, we consider the second order reaction, and also take into account that $[B]=P(x,y,t)$.
Therefore, we have
\[\frac{d\,P}{d\,t}=-kP\cdot [A]\]
with the the reaction rate $k$ and initial condition $[B(t=0)]\equiv P(t=0)=P_0,~[A(t=0)]=[A]_0=N,~[C]_0=0$.

Let us express $[A]$ by $P$. We have
\begin{eqnarray*}
& P\equiv [B]=[B]_0-[C]=P_0-[C]\, , \\
& [A]=[A]_0-[C]= [A]_0-P_0+P\, .
\end{eqnarray*}
Therefore the reaction equation reads
\begin{equation}\label{e_react2}
\frac{d\,P}{d\,t}=-k(N-P_0+P)P\, .
\end{equation}
If during the reaction time, $N-P_0\gg P$ and $N\gg P_0$, the second order reaction equation (\ref{e_react2}) becomes of the first order
\begin{equation}\label{e_react1}
\frac{d\,P}{d\,t}=-kNP\, .
\end{equation}
Substituting Eq. (\ref{e_react1}) in Eq. (\ref{comb_r}) and denoting $C=\rho\cdot k\cdot N$ yields
\begin{equation}\label{comb_react1_a}
\partial_tP=Dh_b\delta(y)\partial_x^2P+ d_{\rho}\partial_y^2P- CP\, .
\end{equation}

\section{Solution of linear reaction-transport comb equation: extinction dynamics}

The reaction term disappears from Eq. (\ref{comb_react1_a}) by substitution
\[P=e^{-Ct}\bar{P}\, .\]
The next step is the Laplace transform $\hat{\cal{L}}[\bar{P}](t)=\tilde{P}(s)$.
This yields
\begin{equation}\label{comb_react1_b}
s\tilde{P}=Dh_b\delta(y)\partial_x^2\tilde{P}+d_{\rho}\partial_y^2\tilde{P}+P_0\delta(y)\, ,
\end{equation}
and the solution can be considered as a product $\tilde{P}(x,y,s)=\tilde{n}(y,s)f(x,s)$
The solution inside fingers is found from the equation in the Laplace domain
\[s\tilde{n}(y)=d_{\rho}\partial_y^2\tilde{n}(y)\]
and the boundary condition $\partial_y\tilde{n}(y)|_{y=\pm h}=0$, which yields
\[\tilde{n}(y,s)=
\frac{\cosh\Big[(h-|y|)\sqrt{s/d_{\rho}}\Big]}{\cosh\Big[(h\sqrt{s/d_{\rho}}\Big]}\]
and $n(y=0)=1$.

The solution is looked for in the form
\begin{equation}\label{solution}
\tilde{P}(x,y,s)=\tilde{n}(y,s)f(x,s)\, ,
\end{equation}
that yields for the second derivative over $y$
\[\partial_y^2\tilde{P}=\Big\{-2\delta(y)\sqrt{s/d_{\rho}}
\frac{\sinh\Big[(h-|y|)\sqrt{s/d_{\rho}}\Big]}{\cosh\Big[(h\sqrt{s/d_{\rho}}\Big]}
+[s/d_{\rho}]\tilde{n}(y)\Big\}f(x,s)\, .\]
Therefore extinct diffusion of magnesium in the balk, determined by $f(x,s)$, is described by equation
\begin{equation}\label{bulk_dif1}
Dh_b\partial_x^2f-2\sqrt{sd_{\rho}}\tanh\Big[h\sqrt{s/d_{\rho}}\Big]f+P_0=0.
\end{equation}
The presence of the hyperbolic tanh distinguishes two time scales/regioons.  At a short time scale, when
$h\sqrt{s/d_{\rho}}\gg 1$,
we have $\tanh[h\sqrt{s/d_{\rho}}]\approx 1$ that yields
\begin{equation}\label{bulk_dif2}
D_{\frac{1}{2}}\partial_x^2f-\sqrt{s}f+f_0=0\, ,
\end{equation}
where $$D_{\frac{1}{2}}=\frac{Dh_b}{2\sqrt{d_{\rho}}}=\frac{D\sqrt{aL}}{2\sqrt{d}}$$
is a generalized diffusion coefficient and $f_0(x)=P_0(x)=
P_0\delta(x)$, for simplicity. \textit{It should be admitted that the generalized diffusion coefficient
in the $x$ direction is independent of the finger's geometry parameters $h$ and $b$.}

Performing the Fourier transform $\bar{f}(k)=\hat{\cal F}[f(x)]$, one obtains
\begin{equation}\label{bulk_dif3}
\bar{f}(k,s)=\frac{P_0}{\sqrt{s}+D_{\frac{1}{2}}k^2}\, .
\end{equation}
Performing the Laplace inversion of the solution (\ref{bulk_dif3}) and taking into account a definition of the Mittag-Leffler function \cite{bateman}
\[ E_{\alpha,\beta}(z)=\frac{1}{2\pi i}\int_{G}\frac{r^{\alpha-\beta}e^r}{r^{\alpha}-z}dr\, ,~~~~\alpha,\beta>0\, , \]
one obtains
\begin{equation}\label{bulk_dif4}
\bar{f}(k,t)=\frac{P_0}{2\pi i}\int_{\sigma-i\infty}^{\sigma+i\infty}\frac{e^{st}ds}{\sqrt{s}+D_{\frac{1}{2}}k^2}=
\frac{P_0}{\sqrt{t}}E_{\frac{1}{2}\, ,\frac{1}{2}}\left(-D_{\frac{1}{2}}k^2t^{\frac{1}{2}}\right)\, .
\end{equation}
This solution is valid for the ``short'' time-scale $ t \ll \frac{hLa}{bd}=\frac{h^2}{d_{\rho}} $. Therefore,
the argument of the Mittag-Leffler function is small, and the latter reads \cite{bateman,klafter}
\[ E_{\frac{1}{2}\, ,\frac{1}{2}}\left(-D_{\frac{1}{2}}k^2t^{\frac{1}{2}}\right)\approx
\exp\left[-\frac{D_{\frac{1}{2}}k^2t^{\frac{1}{2}}}{\Gamma(3/2)}\right]\, ,\]
where $\Gamma(3/2)=(1/2)\Gamma(1/2)=\sqrt{\pi}/2$ is the gamma function\footnote{
Note, that this formula is true only for the Mittag-Leffler function $E_{1/2,1}(z)=E_{1/2}(z)$. However, we can use it approximately here, admitting that the denominator in the exponential $\Gamma(3/2)$ is not correct. It should be stressed that this approximation does not affect the following analysis.}.
Now the Fourier inversion can be easily performed that yields
\[\frac{1}{2\pi}\int_{-\infty}^{\infty}e^{-ikx}e^{-ak^2}dk=
\left[\frac{\Gamma(3/2)}{4\pi D_{\frac{1}{2}}t}\right]^{\frac{1}{2}}
\exp\left[-\frac{x^2\Gamma(3/2)}{D_{\frac{1}{2}}\sqrt{t}}\right]\, .\]
Taking into account the reaction term, one obtains the solution for the short-time scale exstinc diffusion in
the bulk
\begin{equation}\label{bulk_dif5}
P(x,y=0,t<t_0)=\frac{P_0e^{-Ct}}{\sqrt{8D_{\frac{1}{2}}\pi t^3}}
\exp\left(-\frac{\sqrt{\pi}x^2}{2D_{\frac{1}{2}}\sqrt{t}}\right)\, .
\end{equation}
Note, that this result is independent of $h$ and $b$ and coincides with the subdiffusive comb limit
$h\rightarrow\infty$.

On a long time-scale the hyperbolic tanh in Eq. (\ref{bulk_dif1}) has a small argument, which yields
\[\tanh[h\sqrt{s/d_{\rho}}]\approx h\sqrt{s/d_{\rho}}\, .\]
This corresponds to an equation for normal diffusion
\begin{equation}\label{bulk_dif6}
D\sqrt{b/h}\partial_x^2f-sf+P_0/h=0\, ~~~~~~~~D\sqrt{b/h}\equiv D_{bh}
\end{equation}
with a well-known solution for normal diffusion. Finally this yields the long-time scale solution in the bulk
\begin{equation}\label{bulk_dif7}
P(x,y=0,t>t_0)=e^{-Ct}\int dx'P_0(x')
\frac{1}{\sqrt{\pi h^2D_{bh} t}}\exp\left(-\frac{(x-x')^2}{4D_{bh}t}\right)\, .
\end{equation}
When $P_0(x)=\delta(x)$, the Green function coincides with the distribution.
In Eqs. (\ref{bulk_dif5}) and (\ref{bulk_dif7}), the transient time parameter $t_0=\frac{hLa}{bd}$ is determined from the geometry of the experiment.

\section{Reaction front propagation in a case of high concentration of magnesium}

In a case when the concentration of magnesium is high enough, the approximation (\ref{e_react1}) is not valid anymore,
and one has to take into account the second order reaction (\ref{e_react2}). Thus the reaction term reads
\begin{equation}\label{nlinCP}
C(P)=CP+C_1P^2\, ~~~~C=\rho\cdot k\cdot (N-P_0)\, , ~~~~C_1=\rho\cdot k\, .
\end{equation}
In this nonlinear case, an exact analytical treatment
of Eq. (\ref{comb_r}) is impossible, and we apply an analytical approximation to find the overall velocity of the reaction-polymerization front propagation without resolving an exact shape of the front, namely without knowledge of the exact distribution function $P(x,y,t)$.

Since we are seeking for the front propagation in the $x$ direction, the exact shape in the $y$ direction is not important, and we consider the distribution as a function of the $x$ coordinate only $P_1(x,t)$.
In other words, the detail dynamics inside
fingers is not important, and we take into account its overall contribution to diffusion in the $x$ axis .
To this end, the $y$ coordinate is integrated out
\begin{equation}\label{P1}
P_1(x,t)=\int_{-h}^hP(x,y,t)dy\, .
\end{equation}
Therefore, this integration in Eq. (\ref{comb_r}) with reaction due to Eq. (\ref{nlinCP}) yields an equation in the unclosed form. First of all, we take into account integration
\[\int_{-h}^h\partial_y^2P(x,y,t)dy=\partial_yP(x,y,t)\Big|_{y=-h}^{y=h}=0\, .\]
This yields the equation in task
 \begin{equation}\label{combP1}
\partial_tP_1(x,t)=h_bD\partial_x^2P(x,y=0,t)-CP_1-C_1\int_{-h}^hP^2(x,y,t)dy\, .
\end{equation}
We have here two problematic terms. The first one is $P(x,y=0,t)$, which will be expressed by $P_1(x,t)$.
To this end we use the relation (\ref{solution}), where we pay attention on $\tilde{P}(x,y=0,s)=f(x,s)$.
Integrating Eq. (\ref{solution}) over $y$ yields
\[ \tilde{P}_1(x,s)=\tilde{P}(x,y=0,s)\int_{-h}^{h}\tilde{n}(y,s)dy=
2\tilde{P}(x,y=0,s)\int_0^h\tilde{n}(y,s)dy\, .\]
Therefore, one obtains
\begin{equation}\label{PtoP1_a}
\tilde{P}(x,y=0,s)=\sqrt{s/4d_{\rho}}\tanh^{-1}\Big[h\sqrt{s/d_{\rho}}\Big]\cdot  \tilde{P}_1\, .
\end{equation}
Performing the inverse Laplace transform, one obtains from Eq. (\ref{PtoP1_a})
\begin{eqnarray}\label{PtoP1_b}
P(x,y=0,t)&=&\frac{1}{2\pi i}\int_{\sigma-i\infty}^{\sigma+i\infty}ds\int_{0}^{\infty}dt_1dt_2P_1(x,t_1)
\mathcal{R}_1(t_2)\cdot e^{-st_1}\cdot e^{-st_2}\cdot e^{st} \nonumber \\
&=&\int_0^{\infty}dt_1\int_{0}^{\infty}dt_2P_1(t_1)\mathcal{R}_1(t_2)\delta(t-t_1-t_2)=
\int_{0}^{t}dt'P_1(t-t')\mathcal{R}_1(t')\, .
\end{eqnarray}
Here we take into account the causality principle, saying that $P(t-t_1)=0$ for $t_1>t$ and
introduce the kernel $\mathcal{R}_1(t)$ through the Laplace inversion
\begin{equation}\label{kernelR1}
\mathcal{R}_1(t)=\hat{\mathcal{L}}^{-1}\left[\sqrt{s/4d_{\rho}}
\tanh^{-1}\Big(h\sqrt{s/d_{\rho}}\Big)\right]\, .
\end{equation}

The second problematic term is integration of the nonlinear reaction. We present it in a form convenient for
the further analytical treatment. Therefore, we have the following chain of transformations:
\begin{eqnarray*}
N.L.R.T.&\equiv &\int_{-h}^{h}P^2(x,y,t)dy=\frac{1}{(2\pi i)^2}
\int_{\sigma-i\infty}^{\sigma+i\infty}e^{s_1t}ds_1e^{s_2t}ds_2
\int_{-h}^h\tilde{P}(x,y,s_1)\tilde{P}(x,y,s_2)dy=  \\
& =&\frac{1}{(2\pi i)^2}\int_{\sigma-i\infty}^{\sigma+i\infty}e^{s_1t}ds_1e^{s_2t}ds_2\tilde{P}(x,y=0,s_1)
\tilde{P}(x,y=0,s_2)\int_{-h}^h\tilde{n}(y,s_1)\tilde{n}(s_2,y)dy = \\
&=& \frac{1}{(2\pi i)^2}\int_{\sigma-i\infty}^{\sigma+i\infty}\tilde{\mathcal{R}}(s_1,s_2)\tilde{P}_1(x,s_1)
\tilde{P}_1(x,s_2)\frac{\sqrt{s_1s_2}}{4d_{\rho}}\cdot
\tanh^{-1}\Big[h\sqrt{s_1/d_{\rho}}\Big]\cdot
\tanh^{-1}\Big[h\sqrt{s_2/d_{\rho}}\Big] \\
&\times&
e^{s_1t}\,e^{s_2t}ds_1\,ds_2=  \int_{0}^{\infty}d\tau_1d\tau_2P_1(x,\tau_1)P_1(x,\tau_2) \\
&\times&\frac{1}{(2\pi i)^2}\int_{\sigma-i\infty}^{\sigma+i\infty}\tilde{\mathcal{R}}(s_1,s_2)\cdot
\frac{\sqrt{s_1s_2}}{4d_{\rho}}\cdot
\tanh^{-1}\Big[h\sqrt{s_1/d_{\rho}}\Big]\times\tanh^{-1}\Big[h\sqrt{s_2/d_{\rho}}\Big]\cdot
e^{s_1(t-\tau_1)}\,e^{s_2(t-\tau_2)}ds_1\,ds_2 \, ,
\end{eqnarray*}
where
\[\mathcal{R}(s_1,s_2)=\int_{-h}^h\tilde{n}(y,s_1)\tilde{n}(s_2,y)dy\, .\]
Introducing new variables $t-\tau_1=t_1$ and $t-\tau_2=t_2$ yields
\[\int_0^{\infty}d\tau_1=-\int_t^{-\infty}dt_1=\int_{-\infty}^tdt_1\]
and the same for $\tau_2$. Therefore, we have
\begin{equation}\label{nlrt}
N.L.R.T.=\int_{0}^{\infty}dt_1dt_2P_1(x,t-t_1)P_1(x,t-t_2)\cdot
\frac{1}{(2\pi i)^2}\int_{\sigma-i\infty}^{\sigma+i\infty}\tilde{\mathcal{R}}(s_1,s_2)
e^{s_1t_1}\,e^{s_2t_2}ds_1\,ds_2 \, .
\end{equation}
In what follow we will keep this form as the N.L.R.T.

Eventually, equation for the $F_1(x,t)$ reads
\begin{equation}\label{EqFront}
\partial_tP_1=hD\partial_x^2\int_0^t\mathcal{R}_1(t')P_1(t-t')dt'-CP_1-
C_1\int_{-\infty}^t\mathcal{R}(t',t^{\prime\prime})P_1(t-t')
P_1(t-t^{\prime\prime})dt'dt^{\prime\prime}\, ,
\end{equation}
where two-time-point kernel $\mathcal{R}(t',t^{\prime\prime})$ is defined in the Laplace space in Eq. (\ref{nlrt}).

\section{Hyperbolic scaling for the overall velocity of the reaction front propagation}

To evaluate the overall velocity of the asymptotic front,
we follow the hyperbolic scaling consideration, developed in \cite{Fr,F} and adopted in \cite{iomend}
for consideration of the reaction transport front propagation in comb structures. Let us
introduce a small parameter, say $\varepsilon$, at the derivatives
with respect to time and space \cite{Fr,F}. To this end we
re-scale
$$x\rightarrow x/\varepsilon~~ \mbox{and} ~~~ t\rightarrow
t/\varepsilon\, , \mbox{and}~~~P_1(x,t)\rightarrow
P_1^{\varepsilon}(x,t)=P_1\left(\frac{x}{\varepsilon},
\frac{t}{\varepsilon}\right)$$. Therefore, one looks for the
asymptotic solution in the form of the Green's approximation
\begin{equation}\label{WKB}
P_1^{\varepsilon}(x,t)=
\exp\left[-\frac{G^{\varepsilon}(x,t)}{\varepsilon}\right]\, .
\end{equation}
The main strategy of implication of this construction is the limit
$\varepsilon\rightarrow 0$ that yields an asymptotic behavior at finite $x$ and $t$, where
we have
\[\exp\left[-\frac{G^{\varepsilon}(x,t)}{\varepsilon}\right]=0\, ,\]
except for the condition when
\[G^{\varepsilon}(x,t)=0\, .\]

This equation determines the position of the reaction spreading front, and
in this limit,
$G(x,t)=\lim_{\varepsilon\to 0}G^{\varepsilon}(x,t)$ is accounted as the principal Hamiltonian function
\cite{Fr,F}. Therefore, the Hamiltonian
approach can be applied to calculate the propagation front velocity. In
this case partial derivatives of $G(x,t)$ with respect to time and
coordinate have the physical senses of the Hamiltonian and the momentum:
\begin{equation}\label{Action}
\frac{\partial G(x,t)}{\partial t}=-H\, ,~~~~~~~~~
\frac{\partial G(x,t)}{\partial x}=p\, .
\end{equation}

Now, ansatz (\ref{WKB}) for the
probability distribution function inside the bulk is inserted
in Eq. (\ref{EqFront}), where we also make the scaling change
$x\rightarrow\frac{x}{\varepsilon}$ and $t\rightarrow\frac{t}{\varepsilon}$.

Let us start from the last term
in Eq. (\ref{EqFront}), which is the reaction term.
First of all we take into account the following change
in the upper limit of integrations
\[\int_{-\infty}^t dt'\Rightarrow\lim_{\varepsilon\to 0}\int_{-\infty}^{\frac{t}{\varepsilon}}dt'
=\int_{-\infty}^{\infty}dt'\, .\]
Then, we make the following expansion for
$P_1^{\varepsilon}(x,\frac{t}{\varepsilon}-t')\equiv P_1^{\varepsilon}(t-\varepsilon t')$,
which reads
\[P_1^{\varepsilon}(t-\varepsilon t')=\exp\left[-\frac{1}{\varepsilon}G^{\varepsilon}(t-\varepsilon t')\right]\approx
\exp\left[-\frac{1}{\varepsilon}G^{\varepsilon}(t)+t'\partial_tG^{\varepsilon}(t)\right]\, .\]
Therefore, the nonlinear reaction term reads
\begin{equation}\label{hyperbolic1}
N.L.R.T. =e^{-2\frac{G^{\varepsilon}}{\varepsilon}}\cdot\int_{-\infty}^{\infty}dt^{\prime}dt^{\prime\prime}
\frac{1}{(2\pi i)^2}\int_{\sigma-i\infty}^{\sigma+i\infty}\tilde{\mathcal{R}}(s^{\prime},s^{\prime\prime})
e^{-(H-s^{\prime})t^{\prime}}\,
e^{-(H-s^{\prime\prime})t^{\prime\prime}}ds^{\prime}ds^{\prime\prime}\, ,
\end{equation}
where we use the first equation in (\ref{Action}). Integration over $s^{\prime}$ and $s^{\prime\prime}$
is performed with some care, since $\tilde{\mathcal{R}}$
is singular at $s^{\prime}=\pm s^{\prime\prime}$. The main result here is that $\tilde{\mathcal{R}}$ is finite and
$N.L.R.T.\sim  e^{-2\frac{G^{\varepsilon}}{\varepsilon}}$, that in the limit $\varepsilon\rightarrow 0$ is
of the order of $o\left(e^{-\frac{G^{\varepsilon}}{\varepsilon}}\right)$ and, as will be seen in this limit,
does not contribute to the finite result, since the rest of the equation is of the order of
$O\Big(e^{-\frac{G^{\varepsilon}}{\varepsilon}}\Big)$.

Now we consider the the kinetic term taking the time integration with $\mathcal{R}_1$ kernel. This reads
\begin{eqnarray*}
\int_0^{\frac{t}{\varepsilon}}\mathcal{R}_1(t^{\prime})
&{}& e^{-\frac{G^{\varepsilon}(t-\varepsilon t^{\prime},x)}{\varepsilon}}dt^{\prime}=
e^{-\frac{G^{\varepsilon}(t)}{\varepsilon}}
\int_0^{\infty}\mathcal{R}_1(t^{\prime})e^{-Ht^{\prime}}dt^{\prime}\\
&=&
e^{-\frac{G^{\varepsilon}(t)}{\varepsilon}}\tilde{\mathcal{R}}_1(H)\, ,
\end{eqnarray*}
where $\tilde{\mathcal{R}}_1(H)$ is defined in Eq. (\ref{kernelR1}).

Finally, differentiating in the limit $\varepsilon\rightarrow 0$ where  ($N.L.R.T.=0$) and taking
into account that the Hamiltonian $H$ and the momentum $p$ in Eq.
(\ref{Action}) are independent of $x$ and $t$ explicitly (which
leads to the absence of mixed derivatives), one obtains that the kinetic equation
(\ref{EqFront}) becomes a  Hamilton-Jacobi equation:
\begin{equation}\label{HJE}
-\partial_tG=h_bD\Big(\partial_xG\Big)^2\tilde{\mathcal{R}}_1(H)-C\, ,
\end{equation}
where $G\equiv G(x,t)=\lim_{\varepsilon\to 0}G^{\varepsilon}$ is the action (principal Hamiltonian function)
\[G(x,t)=\int_0^t[p(\tau)\dot{x}(\tau)-H(p(\tau),x(\tau))]d\tau\, .\]
The rate $v$ at
which the front moves is determined at the condition $G(x,t)=0$.
Together with the Hamilton equations, this yields
\begin{equation}\label{frontV}
v=\dot{x}=\frac{\partial H}{\partial p}\, ,~~~v=\frac{H}{p}\, . %
\end{equation}
The first equation in (\ref{frontV}) reflects the
dispersion condition, while the second one is a result of the
asymptotically free particle dynamics, when the action is
$G(x,t)=px-Ht$. Taking into account $x=vt$, one obtains Eq.
(\ref{frontV}). The combination of these two equations can be
replaced by
\begin{equation}\label{minmin}
v=\min _{H>0}\frac{H}{p(H)}=\min _{p>0}\frac{H(p)}{p}.
\end{equation}
To proceed, we consider two time limits of the kernel/dispersion term $\tilde{\mathcal{R}}_1(H)$ in Eq. (\ref{HJE}).

\subsection{short time limit $h\sqrt{H/d_{\rho}}\gg 1$}

In the short time limit, which corresponds to subdiffsion on the intermediate asymptotic times,
we have $\tanh\Big[h\sqrt{H/d_{\rho}}\Big]=1$ that yields $ \tilde{\mathcal{R}}_1(H)\approx \sqrt{H/4d_{\rho}}$.
Taking this into account, one obtains from Eq. (\ref{HJE})
\begin{equation}\label{Vshort1}
\frac{(H+C)^2}{H}=\frac{(h_bD)^2}{4d_{\rho}}p^4\, ,
\end{equation}
and $p(H)=\frac{(4d_{\rho})^{\frac{1}{4}}}{\sqrt{hD}}\cdot\frac{\sqrt{H+C}}{H^{\frac{1}{4}}}$.
Taking logarithm from Eq. (\ref{Vshort1}) and differentiation over $p$, one obtains
\[\frac{2}{H+C}=\frac{5}{H}\, .\]
This equation has no solution for $C>0$. Therefore, for $C>0$, $v=\min _{H>0}\frac{H}{p(H)}=0$.
This means the failure of the front propagation, as expected for the initial concentration of magnesium less than actin.

For $C<0$ that corresponds $P_0>N$, one obtains the solution $H=5|C|/3$ that yields a nonzero velocity of the reaction
front propagation at subdiffusion at the initially short times
  \[v=\Big(\frac{5}{3}\Big)^{\frac{5}{4}}\cdot\Big(\frac{2}{3}\Big)^{\frac{1}{2}}
\Big(\frac{D^2h_b^2|C|^3}{4d_{\rho}}\Big)^{\frac{1}{4}}=
\Big(\frac{5}{3}\Big)^{\frac{5}{4}}\cdot\Big(\frac{2}{3}\Big)^{\frac{1}{2}}
\Big(\frac{DaL|C|^3}{4d}\Big)^{\frac{1}{4}} \, .\]
Again, we admit that this result is independent of the finger geometry and correspondingly, has the comb limit
$h\rightarrow\infty$.

\subsection{large time limit $h\sqrt{H/d_{\rho}}\ll 1$}
In the opposite case of the large time asymptotics, $\tanh\Big[h\sqrt{H/d_{\rho}}\Big]\approx h\sqrt{H/d_{\rho}}$,
and $\tilde{\mathcal{R}}_1(H)\approx\frac{1}{2h}$. Therefore, as seen from  Eqs. (\ref{bulk_dif6}) and (\ref{bulk_dif7}), we have from Eq. (\ref{HJE})
\begin{equation}\label{Vlong1}
H=\frac{D_{bh}}{2}p^2-C\, ,
\end{equation}
and $p=\sqrt{2(H+C)/D_{bh}}$. This immediately yields
\[v=\min_{H>0}\Big[\frac{HD_{bh}^{\frac{1}{2}}}{\sqrt{2(H+C)}}\Big]=0 \]
for $H=0$, as expected for $C>0$.

The situation changes dramatically for $C<0$ ($P_0>N)$. In this case, the equation
\[\frac{\partial H}{\partial p}=\frac{H}{p}\]
yields
\[2\bar{D}p=\bar{D}p+\frac{|C|}{p}\\, .\]
Here we scale the diffusion coefficient $\bar{D}=\frac{D_{bh}}{2}$.
This yields the solution for $p=\sqrt{|C|/\bar{D}}$. Therefore, substitution in the velocity equation yields
\[v=\min_{p>0}\frac{H(p)}{p}=\frac{\bar{D}p^2+|C|}{p}=2\sqrt{\bar{D}|C|}\, .\]
This is a well known Fisher-KPP result for the reaction front velocity with the scaled diffusivity $\bar{D}$.

\section{Conclusion}

In the paper we considered anomalous transport and reaction dynamics by providing
the theoretical grounds for the possible experimental realization of actin polymerization in 
comb-like geometry. Different regimes for the reaction of polymerization have been 
considered, and depending on the concentration
of reagents (magnesium and actin), we recovered both the failure of reaction front propagation and a finite speed corresponding to the Fisher-KPP long time asymptotic regime.
Finally, we repeat that our analytical results on reaction propagation can help to guide
the design of microfluidic devices but also can lead to real
experimental tests of anomalous diffusion and reaction dynamics.

This research was supported by the Israel Science Foundation (ISF).


\begin{thebibliography}{9}

\bibitem{pfohl1} S. Köster, J. Kierfeld, and  T. Pfohl,  Eur. Phys. J. E, \textbf{25}, 439
(2008).

\bibitem{pfohl2} S. Deshpande and T. Pfohl, Biomicrofluidics, \textbf{6},  034120 (2012).

\bibitem{white}
S.R. White and M. Barma, { J. Phys. A: Math. Gen.} \textbf{17}, 2995 (1984).

\bibitem{weiss}
G.H. Weiss and S. Havlin, { Physica A} \textbf{134}, 474 (1986).

\bibitem{havlin}
O. Matan, S. Havlin, and D. Staufler, {J. Phys. A: Math. Gen.} \textbf{22}, 2867 (1989).

\bibitem{iomend} A. Iomin and V. M\'endez, Phys. Rev. E. \textbf{88}, 012706 (2013).

\bibitem{sokolov} D. Froemberg, H. Schmidt-Martens, I.M. Sokolov, and
F. Sagu´es, Phys. Rev. E \textbf{78}, 011128 (2008); \textbf{83}, 031101 (2011).

\bibitem{iomsok} A. Iomin and I.M. Sokolov, Phys. Rev. E \textbf{86}, 022101 (2012).

\bibitem{kpp} A. Kolmogoroff, I. Petrovskii, and N. Piscounoff, Mosc. Univ.
Math. Bull. \textbf{1}, 25 (1937).

\bibitem{fisher} R.A. Fisher, Annals of Eugenics \textbf{7}, 355 (1937).

\bibitem{dkm2004} A.M. Dykhne, P.S. Kondratenko, and L.V. Matveev,
JETP Lett. \textbf{80}, 410 (2004).

\bibitem{dvork2009} O.A. Dvoretskaya and P.S. Kondratenko, 
Phys. Rev. E \textbf{79}, 041128 (2009).

\bibitem{dvorkm2010} O.A. Dvoretskaya, P.S. Kondratenko, and L.V. Matveev, 
JETP \textbf{110}, 58 (2010).


\bibitem{bateman} H. Bateman and A. Erd\`elyi, \textit{Higher transcendental functions} (McGraw-Hill, NY, 1955).

\bibitem{klafter} R. Metzler and J. Klafter, Phys. Rep. \textbf{339}, 1 (2000).

\bibitem{Fr} M. Freidlin, \textit{Markov Processes and Differential
Equations: Asymptotic Problems } (Birkhauser, Basel, 1996).

\bibitem{F} S. Fedotov, Phys. Rev. Lett. \textbf{86}, 926 (2001).




\end{thebibliography}
\end{document}